\documentclass[manuscript]{aastex}

\slugcomment{To appear in the Astrophysical Journal}

\shorttitle{Mid-Infrared Imaging of the Bipolar Planetary Nebula M2-9 from \emph{SOFIA}}

\shortauthors{Werner et al.}

\begin{document}

\title{Mid-Infrared Imaging of the Bipolar Planetary Nebula M2-9 from \emph{SOFIA}}

\author{M. W. Werner\altaffilmark{1}, R. Sahai\altaffilmark{1}, J. Davis \altaffilmark{1}, J. Livingston \altaffilmark{1}, F. Lykou \altaffilmark{2}, J. de Buizer \altaffilmark{3}, M. R. Morris \altaffilmark{4}, L. Keller \altaffilmark{5}, J. Adams \altaffilmark{6}, G. Gull \altaffilmark{6}, C. Henderson \altaffilmark{6}, T. Herter \altaffilmark{6}, J. Schoenwald \altaffilmark{6}}

\altaffiltext{1}{Jet Propulsion Laboratory, California Institute of Technology, 4800 Oak Grove Drive, Pasadena, CA 91107 USA.  Michael.W.Werner@jpl.nasa.gov}
\altaffiltext{2}{Institute for Astronomy, University of Vienna, Turkenschanzstrasse 17, A-1180, Vienna, Austria}
\altaffiltext{3}{USRA  SOFIA Science Center, M/S 211-3, NASA Ames Research Center, Moffett Field, CA 94035 USA}
\altaffiltext{4}{Division of Astronomy, PO Box 951547, UCLA, Los Angeles, CA 90095 USA}
\altaffiltext{5}{Department of Physics, Ithaca College, Ithaca, NY 14850  USA}
\altaffiltext{6}{Department of Astronomy, Cornell University, Ithaca, NY 14853 USA}

\begin{abstract}
We have imaged the bipolar planetary nebula M2-9 using \emph{SOFIA}'s FORCAST instrument in six wavelength bands between 6.6 and 37.1 $\mu m$. A bright central point source, unresolved with \emph{SOFIA}'s $\sim$ 4${''}$-to-5${''}$ beam, is seen at each wavelength, and the extended bipolar lobes are clearly seen at 19.7 $\mu m$ and beyond. The photometry between 10 and 25 $\mu m$ is well fit by the emission predicted from a stratified disk seen at large inclination, as has been proposed for this source by Lykou et al and by Smith and Gehrz. The principal new results in this paper relate to the distribution and properties of the dust that emits the infrared radiation. In particular, a considerable fraction of this material is spread uniformly through the lobes, although the dust density does increase at the sharp outer edge seen in higher resolution optical images of M2-9. The dust grain population in the lobes shows that small ($<$ 0.1 $\mu m$) and large ($>$ 1 $\mu m$) particles appear to be present in roughly equal amounts by mass. We suggest that collisional processing within the bipolar outflow plays an important role in establishing the particle size distribution.
\end{abstract}

\keywords{planetary nebulae: individual (M2-9)}

\section{INTRODUCTION}
Although planetary nebulae (PNs) evolve from (initially) spherically--symmetric mass-loss envelopes around AGB stars, modern ground-based and Hubble Space Telescope (HST) imaging surveys have shown that the vast majority of PNs deviate strongly from spherical symmetry (e.g., Schwarz, Corradi, \& Melnick 1992, Sahai, Morris, \& Villar 2011b).
The morphologically unbiased survey of young PNs with HST (Sahai \& Trauger 1998, Sahai et al. 2011b) shows that PNs with bipolar and multipolar morphologies represent almost half of all PNs, as was previously pointed out by Zuckerman and Aller (1986). The significant changes in the circumstellar envelope morphology during the evolutionary transition from the AGB to the PN phase require a primary physical agent or agents which can break the spherical symmetry of the radiatively-driven, dusty mass-loss phase. Sahai and Trauger (1998) proposed that the primary agent for breaking spherical symmetry is a jet or collimated, fast wind (CFW) operating during the early post-AGB or late AGB evolutionary phase. The nature of the central engine that can produce such CFW's is poorly understood, although a number of theoretical models, most of them requiring the central star to be a binary, have been considered (Morris, 1987; Garcia-Segura, 1997; Balick \& Frank, 2002; Matt, Frank \& Blackman, 2006). Detailed multi-wavelength studies of individual bipolar and multipolar PNs that can constrain the physical properties of the lobes produced by the CFW's and the central regions are needed to test these models.
\\

M2-9 is a well-studied bipolar nebula at many wavelengths from the optical (e.g., Solf 2000) to the radio (Kwok et al. 1985), and is usually classified as a planetary nebula, although it has also been suggested that this object has a symbiotic star at its center (e.g., Schmeja \& Kimeswerner 2001), or belongs to the compact planetary nebula (cPNB[e]) subclass of B[e] stars (Frew \& Parker 2010). Each of the bipolar lobes appears in optical, emission-line images as a collimated, limb-brightened structure with an ``inner" lobe of radial extent about $29{''}$ and width $12{''}$, and a much fainter ``outer" lobe of radial extent about $62{''}$ and similar narrow width (e.g., Corradi, Balick, \& Santander-Garcia 2011, hereafter Co11). Proper motion of bright ansae at the tips of the faint lobes implies a radial expansion speed of 147 $km\,s^{-1}$ (Co11). One of the most striking phenomena observed in M2-9 is a pattern of emission-line knots in each lobe that appears to rotate with a period of 90\,yr, interpreted as resulting from either a pair of rotating light-beams (Livio \& Soker 2001) or jets (e.g., Co11). Livio \& Soker (2001) propose a model for producing the light-beams in which jets clear a path which allows ionizing radiation from a white-dwarf companion of the primary AGB (or post-AGB) star, to irradiate the knot regions.
\\

The dense, dusty waist separating the two lobes was first mapped with a $3{''}\times5{''}$ beam in the CO(J=2--1) line with the Plateau de Bure millimeter-wave interferometer (PdBI) by Zweigle et al. (1997) revealing the presence of a large ($\sim\,6{''}$ diameter) ring structure with a mass of $\sim$ 0.01$M_{\odot}$. More recent PdBI mapping with a $0\farcs8\times0\farcs4$ beam reveals a second inner ring that is almost three times smaller than the outer one (Castro-Carrizo et al. 2012). The rings are co-planar and seen almost edge-on, with their axes being inclined at $\sim\,19\arcdeg$ to the sky-plane, similar to the inclination of the axis of the bipolar lobes (Solf 2000). 

\section{PREVIOUS INFRARED STUDIES OF M2-9}

The mid-infrared observations from \emph{SOFIA} reported here build on previous studies of M2-9 by Smith and Gehrz (2005; hereafter SG05) and Lykou et al. (2011; hereafter Lyk11). SG05 imaged M2-9 at 8.8, 17.9, and 24.5 $\mu m$, using the \emph{IRTF}. Their results for the flux from the central point source, using a smaller effective aperture, are consistent with ours. They also detect the lobes at all three wavelengths, although the images they present of the extended emission are less extensive and of lower signal-to-noise than the present results. More recently, Lyk11 present an extensive study of M2-9, reporting spectroscopy from both $ISO$ and \emph{Spitzer} out to $\sim$ 35 $\mu m$ and summarizing previous measurements as well. They also present ground-based interferometric observations in the 10 $\mu m$ region which identify a disk of dimension $\sim$ 40 $mas$ within the central unresolved point source. We have adapted the model for the central source used by Lyk11 (see also Chesneau et al., 2007) to the analysis of our photometry.  Lagadec et al. (2011) have also published recent photometry of M2-9 in the 8-13 $\mu m$ region, reporting fluxes $\sim$ 20-to-30\% higher than found here or given by Lyk11. This discrepancy may be due in part to the various photometric bands used for the different measurements. Finally, Sanchez-Contreras et al. (1998; hereafter SC98) present an image of M2-9 at 1.3 $mm$ which suggests a significant amount of very cold dust associated with the lobes of the nebula. In addition, M2-9 was measured in the $IRAS$, $AKARI$, and $WISE$ surveys.

\section{OBSERVATIONS}
We observed M2-9 with \emph{SOFIA}'s FORCAST instrument (Herter et al., 2012), which provides imaging capability in multiple spectral bands between 5 and 40 $\mu m$, using two $256\times256$ blocked impurity band detector arrays. For many observations, the arrays are used simultaneously with a dichroic beam splitter. Wavelengths from 5-25 $\mu m$ are directed to a Si:As array, while the 25-40 $\mu m$ wavelengths pass through to a Si:Sb array. After correction for focal plane distortion, FORCAST effectively samples at 0.768 $arcsec\,pixel^{-1}$, which yields a $3.2arcmin\times3.2arcmin$ instantaneous field of view in each camera.
\\

We elected to observe M2-9 in six bands at 6.6, 11.1, 19.7, 24.2, 33.6, and 37.1 $\mu m$. Each of these filters has a bandwidth of $\sim$ 4-to-30\% (see Herter et al. 2012 for further information about the FORCAST instrument). The observations were made on two separate \emph{SOFIA} flights on 11 May and 2 June 2011. The 6.6 and 11.1 $\mu m$ data were taken sequentially with a mirror in place of the dichroic, while the 24.2 $\mu m$ data were taken simultaneously with the 37.1 $\mu m$ data, and the 19.7 $\mu m$ data with the 33.6 $\mu m$ data. The chopping secondary on \emph{SOFIA} was configured to chop east-west with a 30${''}$ amplitude on the sky (perpendicular to the long axis of the nebula, which is very close to north-south) to cancel atmospheric emission. The telescope was nodded east-west every 30 sec with a $30{''}$ throw to facilitate subtraction of (predominantly) telescope radiative offsets. This chopping and nodding strategy made it possible to keep an image of the nebula within the field of view of the array continually during the observations.
\\

The total observation time in each filter for the observations presented here is about 10 minutes, with data being written to a FITS image approximately every second, and with approximately half the total exposure time in each filter coming from each of the two flights.  The data were reduced and calibrated with pipeline software at the \emph{SOFIA} Science Center and a series of FITS files were posted in the \emph{SOFIA} Archive for download by the investigator team.

\section{RESULTS}
\subsection{Images}
In Figure 1 we show the final \emph{SOFIA} images of M2-9 in all six bands. Also shown, for comparison, is a composite line emission image of M2-9 from HST, as well as a continuum image from HST in the 547 $nm$ filter; the HST images were obtained on 1997 Aug 7 with WFPC2 as part of GO program 6502 (PI: B. Balick). In each of the \emph{SOFIA} images, a compact central source is apparent, and the extended emission lobes are clearly seen at 19.7 $\mu m$ and longward. N is to the top and E to the left in these images, so the position angle on the sky of M2-9 is very close to N-S. Therefore, we refer to scans parallel and perpendicular to the outflow-lobes as N-S and E-W scans, respectively. In Figure 2, we show North-South scans at 19.7 and 37.1 $\mu m$ extending more than 20${''}$ from the central compact source. At the longer wavelength, the compact source is superposed on the emission from the lobes, which contribute a much larger fraction of the total flux than at the short wavelengths (see Table 1 and Figure 1). We emphasize that the extended wings due to the lobe emission seen at 37.1 $\mu m$ in Figure 2 are not seen in the point source PSF (cf. Figure 7).

\subsection{Photometry}
Photometry of the central source has been carried out at each wavelength using the point source photometry routine in ATV, which also provides an estimate of the FWHM of the point source. Based on the scans shown in Figure 2, we set the aperture radius for this photometry to be $5.4{''}$ (7 pixels) and the reference sky annulus to be between $5.4{''}$ and 6.9${''}$ (9 pixels). This choice helped us to determine the compact source flux with minimum contamination from the surrounding plateau of emission. These results are tabulated in Table 1.  Also given in Table 1 is the total flux at each wavelength, as determined by integrating the total flux within the $20{''}\times40{''}$ area shown in Figure 1 and subtracting the average sky brightness determined from $20{''}\times40{''}$ areas N, S, E, and W of M2-9 on the images. The third column in the table gives the difference between the compact source flux and the total flux, which is an estimate of the flux in the extended lobes plus any extended component in the EW plane of the compact source. Note that although the lobes are not readily visible in the images at 6.6 and 11.1 $\mu m$, they are detected at these wavelengths in the integrated emission from the source; the flux tabulated for the extended component at these wavelengths in Table 1 is consistent with that which can be estimated for the lobes at 8.8 $\mu m$ from the images presented by SG05.  For completeness, we include fluxes at the longer wavelengths as measured by $IRAS$ and by SC98 as well as the flux measured by ISO shortward of 5 $\mu m$ as reported by Lyk11; the central compact source was not resolved by ISO's 1${''}$ to 2${''}$ beam at these short wavelengths. The point source FWHM reported by ATV varied from 3.7${''}$ at 6.6 and 11.1 $\mu m$ to 4.9${''}$ at 37.1 $\mu m$. At all wavelengths, the observed FWHM agrees with the recommended value for this flight series provided by the \emph{SOFIA} Science Center (Table 1). Thus there is no evidence that \emph{SOFIA} has resolved the central source at any wavelength. 
\\

In Figure 3 we plot the total flux from the $20{''}\times40{''}$ area of Figure 1. Note from Figure 3 that most of the flux measured by $IRAS$ shortward of 60 $\mu m$ comes from this area. The S/N of our \emph{SOFIA} measurements is quite high; for both lobes and point source the principal uncertainty in almost all cases is the $\pm20\%\, (3\sigma)$ calibration uncertainty (Herter, 2012).

\section{ANALYSIS AND INTERPRETATION}
\subsection{Emission Mechanism and Total Luminosity}
The spectral and spatial characteristics of the emission from the lobes are suggestive of emission from dust. Fine structure emission lines appear in \emph{Spitzer} spectra of the lobes (Lyk11), but they are not strong enough to contribute substantial flux to that measured in \emph{SOFIA}'s broad filters: The strongest lines which lie in any of our band passes are [SIII] at 18.7 $\mu m$, [OIV] at 26.4 $\mu m$, and [SiII] at 35 $\mu m$, but comparison of the line intensity with that of the adjacent continuum shows that the lines contribute only 1-to-2\% of the total flux measured with \emph{SOFIA}. The weak PAH emission seen at 11.3 $\mu m$ does not contribute significantly to the integrated flux in the 11.1 $\mu m$ band, and no PAH emission is seen in the ISO spectra centered on the central point source. There is ample evidence from previous studies of scattered light and polarization that the lobes contain dust.  We thus interpret the radiation from both lobes and the point source as being due to emission from dust.
\\

Integrating over the SEDs tabulated in Table 1, and assuming isotropic emission and a distance to the source of 1200 $pc$ (see below), we find that the total 2.5-40 $\mu m$ infrared luminosity of the compact source is $\sim$ 840 $L_{\odot}$, that of the lobes $\sim$ 390 $L_{\odot}$. The total 2.5-120 $\mu m$ luminosity of M2-9, including the $IRAS$ measurements at longer wavelengths, is $\sim$ 1530 $L_{\odot}$.  The observed luminosity at shorter wavelengths is no more than a few percent of that seen in the infrared.

\subsection{Modeling the Central Point Source}
At each wavelength the central point source appears unresolved, with a measured FWHM close to the value recommended by the \emph{SOFIA} Science Center for the flight series during which M2-9 was observed. However, the SED of the central source is much broader than a blackbody, suggesting that a range of dust temperatures is being sampled, as would be the case for an optically thin or somewhat face-on disk-like geometry. This type of geometry has previously been proposed for this central source by SG05, based on similar arguments. Lyk11 report interferometric imaging of a compact $\sim$ $0.037{''}\times0.046{''}$ dust disk at the center of M2-9, and they have produced a model of a circumbinary disk (see Chesneau et al. 2007 for details), suggesting that the interferometric measurements sample the warm inner regions of the disk. This model has now been updated to fit the \emph{SOFIA} data on the central point source. As is shown in Figure 4, the fit is excellent at wavelengths from 11.1 to 24.2 $\mu m$ over which most of the energy from this source is observed.  The parameters for this model are detailed in Table 2. Neither our observations nor those of Lyk11 constrain the disk outer radius.
\\

The distance to M2-9 is quite uncertain, as is often the case for planetary nebulae. 
The present observations do not constrain the distance, so we have adopted D=1200 $pc$ for consistency with Lyk11. This in agreement with the recent careful estimate of $1.3 \pm 0.12$ $kpc$, based on kinematic analysis of motions of features in the lobes (Co11). 
We emphasize, however, that the principal new results of this work, which relate to the spatial distribution and particle size distribution of the dust, are derived directly from the observations and are independent of the adopted distance.
\\

At a distance of 1200 $pc$, the angular extent of the disk modeled in Figure 4 is less than 2${''}$. Thus there is ample room for cooler material exterior to this disk, perhaps associated with the inner CO-emitting disk most recently discussed by Castro-Carrizo et al. (2012), which could produce the radiation seen at 33.6 and 37.1 $\mu m$ in excess of the model prediction without producing a spatially resolved source at these wavelengths. The flux measured with \emph{SOFIA} at 6.6 $\mu m$ and by $ISO$ from 2.5-to-6 $\mu m$ is also in excess of the predictions of the model, and may be due to additional scattered or thermal emission leaking outwards from warm dust close to the star that resides in a different geometrical component than the disk. A plausible origin of this component may be a dusty wind from the disk (see 5.3.2).
\\

There is no discrepancy between the observed luminosity, 1530 $L_{\odot}$, and the modeled stellar luminosity, 2500 $L_{\odot}$. As shown by the bipolar geometry, this object is markedly asymmetric. It is likely that more power emerges perpendicular to the disk, that is, in the plane of the sky along the general direction of the outflow, than is radiated into our direction. In addition, it appears that the lobes may not be optically thick to the heating radiation (see below). These facts could account for the (less than a factor of two) difference between the two luminosity estimates.  
\\

\subsection{Extended Emission}
\subsubsection{Particle Size and Composition}
The SED and temperature of the emission from the lobes shows that the dust particles are considerably smaller than the heating wavelengths around 1 $\mu m$. This is apparent from Figure 5 and Table 1, which show that the emission from the lobes peaks at a wavelength of 35-to-40 $\mu m$, corresponding to a grain temperature of around 100 $K$. A black particle at a projected distance of 10${''}$ from the point source (assumed to have a luminosity of 2500 $L_{\odot}$ and to be 1200 $pc$ from Earth) would have a temperature of around 20 $K$. Therefore, the particles in the lobes that produce the emission seen by \emph{SOFIA} at 19.7 $\mu m$ and beyond have to be small.
\\

Based on this observation we have fit the SED of the extended source of M2-9, defined as the total observed flux minus that of the point source. 
We have used the DUSTY spherically-symmetric dust radiative transfer code (Ivezic et al. 1999) to fit the SED longward of $\sim$ 20 $\mu m$. We assumed a central illuminating blackbody with an effective temperature $T_b=5000\,K$ and luminosity $L\sim\,2500\,L_{\odot}$ surrounded by a shell with an $r^{-2}$ radial-density distribution of $\sim$ 0.1 $\mu m$ amorphous carbon particles (dust-type amC in the code). We justify the use of carbon dust, rather than the (oxygen-rich) silicate dust used in the Lyk11 model, later in this section. We have roughly accounted for the non-spherical geometry of the emitting region in our modeling as follows. We approximate the emitting region as covering a solid angle $2\pi$ (instead of the $4\pi$ covered by a spherical shell), and therefore scale the model output flux by a factor 0.5 when fitting to the observed fluxes. Our derived model parameters below are not too sensitive to the geometry because they are constrained by the mid- and far-infrared emission, which is optically-thin. The results of the fit are shown in Figure 5; data from \emph{AKARI}, \emph{WISE}, and \emph{IRAS} are used in addition to the \emph{SOFIA} data.
\\

We find that the grains in the shell are warm, with equilibrium temperatures varying from 95 to 40\,K from the inner to outer radius of the shell (i.e., from 5\farcs4 to $20{''}$). The radial optical depth of the shell, in the visible, is $\tau _V=1$. The total dust mass of the shell is estimated using Eqn. 2 of Sarkar \& Sahai (2008), to be 0.001$M_{\odot}$, assuming $\kappa _{60\,\mu m}$=150 cm$^{2}$\,g$^{-1}$ (Jura 1986). The shell mass scales linearly with the outer radius. Our model flux falls increasingly below the observed values for wavelengths longer than $\sim$ 70 $\mu m$. However, simply increasing the outer radius is not adequate for decreasing this discrepancy as shown by models with outer radii $>20{''}$; a population of cooler grains is needed that does not reside in the lobes. We suggest that such grains may reside in the low-latitude regions of the dusty equatorial waist of the nebula, perhaps associated with the molecular rings, and/or beyond their radial extent.
We have also not attempted to fit the lobe flux shortward of 20 $\mu m$, where the model fluxes are considerably less than observed. It is possible that this excess is due to single photon heating as described below.
Consideration of the 24.2/37.1 $\mu m$ flux ratio dictated our choice of using carbon dust. We found that although we can construct models with silicate dust that reproduce the SED of the extended source in M2-9 just as well as those with carbon dust, the silicate-dust models are not able to produce the observed 24.2/37.1 $\mu m$ flux ratio at the inner radius of the lobes - the model ratio is about 0.22, significantly lower than observed [cf. Figure 6]. This is because the silicate grains at this radius are cooler than carbon grains.
\\

Although the model fit to the SED over the range of peak emission from \emph{SOFIA} looks excellent, further exploration shows that the model is not totally adequate. In Figure 6, we show the 24.2/37.1 $\mu m$ flux ratio as a function of position along the midline of the lobes, moving northwards from the central source. Close to the central source, the observed ratio agrees well with the predictions of the model, decreasing as expected with distance from the star. However, starting at about 8${''}$ from the source the observed ratio levels off and no further decrease is seen. A similar effect is seen in the 19.7/37.1 $\mu m$ flux ratio. We suggest that this is due to transient heating of the small grains by single photons becoming the dominant heating mechanism in the outer portions of the nebula. This naturally leads to a distance-independent reradiated spectrum. A complete treatment of this idea is beyond the scope of this paper. However, we note that Castelaz, Sellgren, \& Werner (1987) show that transient heating is important out at least to 25 $\mu m$ within a few arc minutes of 23 Tau in the Pleiades, at an assumed distance of 125 $pc$.
This supports our suggestion that we have observed this phenomenon within 10${''}$ of a star that has comparable luminosity but is ten times further away.
\\

While further modeling would be warranted, we note that the models shown in Figures 4 and 5 provide satisfactory fits to the data at the wavelengths where both the compact source and the lobes emit most of their energy as seen from Earth. However, one further point merits discussion, which is the long wavelength emission seen by SC98 at 1.3 $mm$. Our model flux falls far below this measurement at 1.3 $mm$. Although free-free emission is present both towards the center and in the lobes, 
it does not dominate the 1.3 $mm$ flux.
In the model of free-free emission of this object by Kwok et al. (1985), the spectrum of the lobes turns over at $\sim$ 1\,GHz, limiting its contribution to be less than a factor 10 of the measured $mm$-wave flux from the lobes. We conclude, in agreement with SC98, that there is a substantial component of rather cold, large ($>1\,\mu m$) grains in the lobes. 
SC98 estimate that (adjusted to the 1200 $pc$ distance we have adopted for M2-9) the lobes contain $\sim$ 0.0015 $M_{\odot}$ of cold dust particles with radii of
1.5-to-20 $\mu m$ 
if the emission is attributed to amorphous carbon. 
It is noteworthy that this is comparable to the $\sim$ 0.001 $M_{\odot}$ of small particles required to fit the shorter wavelength radiation, as discussed above.
SG05, using the IRAS data at wavelengths $\gtrsim$ 25um, derive a mass for the dust producing the emission from the lobes of $\sim$ 0.005 $M_{\odot}$ for “carbon” grains.  The five-fold discrepancy with our value of $\sim$ 0.001 $M_{\odot}$ is largely due to the fact that they attribute the emission to graphite grains, and adopt a 5$\times$ lower mass absorption coefficient than the 160 $cm^2$ $gm^{-1}$ adopted here for amorphous carbon.
\\

The recent PdBI observations by Castro-Carrizo et al. (2012) with a $0\farcs8\times0\farcs4$ beam reveal an unresolved source of 1.3\,$mm$ continuum emission with flux 240\,mJy associated with the central disk. This is in agreement with the $\sim$ 210 mJy estimated for the central source at this wavelength by SC98. 
By extrapolating the ionized-wind model of Kwok et al. (1985), we estimate that the contribution of the emission from ionized gas to the core flux at 1.3 $mm$ is about 90 $mJy$ or less.  Hence the thermal dust emission from the core is at least 150 $mJy$, and since the (extrapolated) disk model flux at 1.3 $mm$ falls far below this value, we infer that like the lobes, the central region must also contain a substantial population of large grains.  This agrees with the increasing observational evidence for the presence of large grains in the central regions of post-AGB objects (Sahai et al. 2011a).

\subsubsection{Particle Size Distribution}
It is striking that comparable masses of large (radii $>$ 1 $\mu m$) and small (radii $<$ 0.1 $\mu m$) grains are present in the M2-9 lobes.  We speculate that large grains are present in the central disk source (as suggested by the 1.3 $mm$ flux of the central source), and grain-grain collisions between these produce small particles. Both small and large particles are driven out of the disk by radiation pressure by the starlight, forming a disk wind. This mechanism has been proposed by Jura et al. (2001) to explain the far-infrared excesses observed towards the red giant SS Lep.  Sputtering of large grains by shocks due to the interaction of high-velocity outflows with slowly-expanding circumstellar material may further enhance the population of small grains in the lobes. 
The grain size distribution in M2-9 may be far from the equilibrium power law established in a collisional cascade.  This is consistent with the short time scales which characterize this source, which has a dynamical age of $\sim$ 2500 years (Co11).

\subsubsection{Spatial Distribution of the Emission}
In Figure 1 we show an image of M2-9 in the HST 547 $nm$ filter, which samples continuum emission due, presumably, to scattered light. In Figure 7 we compare scans through the Northern lobe at a position 10${''}$ N of the point source at 24.2 and 37.1 $\mu m$ with the corresponding scan through the HST 547 $nm$ image. The emission measured with \emph{SOFIA} shows much less structure and much greater symmetry than seen in the HST image, which shows significant limb brightening, but only on the Eastern edge of the lobe. Thus the thermal emission seen in the mid-infrared is more uniformly distributed than the scattered light seen in the visible.
\\

Although one might expect the mid-infrared emission to be produced by dust located exterior to the lobes, where it might be associated with material that confines the outflow, our results show quite clearly that a good fraction -- or perhaps all -- of the mid-infrared radiation seen by \emph{SOFIA} arises interior to the lobes as traced by optical images.  We adopt a simple model in which the emission arises in an annular cylinder aligned with the observed lobes and convolve the resultant profile with the \emph{SOFIA} beam to compare with the data. 
We do a 1-dimensional convolution in the EW direction. The source is uniform enough in the NS direction to make this an appropriate approach.
We wish to compare an EW scan across the lobe 10${''}$ N of the central point source with the predictions of this model. At this position, the HST 547 $nm$ continuum image shows that the FWZI of the observed lobe is 12${''}$. We take half of this, or 6${''}$, as the radius of the annular cylinder at this position.  We compare the data at 37.1 $\mu m$, averaged over 3 pixels ($\sim$ 2${''}$) in the NS direction to improve the S/N, with the predictions of the model. For simplicity, we neglect any possible temperature dependence of the emitting material, so we are actually modeling the volume emissivity distribution and assuming that it is equivalent to the dust distribution. The use of the data at 37.1 $\mu m$, our longest and therefore least temperature-sensitive wavelength, should make this an acceptable approximation for an initial calculation, particularly if single-photon heating is important at this wavelength. 
\\

The \emph{SOFIA} scans in Figure 7 show two separate peaks along the scan, with a small depression in the middle, particularly at 24.2 $\mu m$ where the resolution is slightly better than at 37.1 $\mu m$. Thus it is obvious that a uniformly filled lobe cannot fit the data; this is shown in Figure 8a, which compares the scan at 37.1 $\mu m$ with the prediction for a uniformly filled lobe. On the other hand, a model in which the dust is confined to the outer regions of the lobe, as might be the case if material is piled up at the interface between the lobe and its exterior environment, also does not fit the data, as is shown in Figure 8b for the case where the dust occupies only the outer 5\% of the lobe.
\\

Figures 8a and 8b together suggest that a simple linear combination of a uniform dust distribution with one which is concentrated towards the edge of the lobe might provide a good fit to the data. This proves to be the case; in fact several such combinations provide an adequate fit because with a lobe width of $\sim$ 12${''}$ and a beam width close to 5${''}$ (cf. Figure 7), we do not have many statistically independent points in the comparison. As one interesting example, we show in Figure 8c a model which combines the distributions shown in Figures 8a and 8b in such a way that 30\% of the material lies in a uniform distribution all the way to the edge of the lobe while an additional 70\% is confined to the outer 5\% of the lobe. Assuming that the dust and gas are well-mixed, this model could be consistent with the limb brightening seen in some of the optical emission lines, as the density in the outer, narrow annulus would be about 20 times that in the central regions. Note, however, that a model in which the outer emission is confined to a narrow annulus exterior to the visible wavelength lobe provides an equally good fit to the data; the implications of having the increased dust density exterior to the visible lobe are substantially different from those of having the increase interior to the lobe. Higher resolution observations, perhaps from JWST, will be required to distinguish between these possibilities. The basic conclusion of this discussion -- that an appreciable fraction of the infrared emission comes from well inside the lobes, implying as well that the dust is similarly distributed -- is well-established, however.

\subsubsection{Comparing Visible and Infrared Images}
\subsubsubsection{Large Scale Morphology}
The connection -- or lack of connection -- between the dust producing the scattered light at visible wavelengths and that producing the infrared radiation is puzzling. Although the limb brightening on the Eastern edge of the lobe at 547 $nm$ (Figure 7c) might be consistent with the model shown in Figure 8c, a similar brightening expected from the dust distribution is not seen on the Western edge; there is no evidence in the symmetrical infrared images for a preferential brightening of the Eastern limb of the lobe.   The time interval between the HST 547 $nm$ image and our SOFIA measurements is about 14 years.  It is possible that the bright region has moved away from the limb in a manner similar to the motion of the features in the emission line images presented by Co11. However, with an overall period $\sim$ 90 years, during this 14 year interval the bright region would have moved (in projection) less than half of the distance to the center line of the lobe.  It should thus be visible near the Eastern edge of the infrared scan if it were as bright relative to the Western half of the lobe in the infrared as it is in the visible. 
\\

The average 547 $nm$ surface brightness of the Western half of the lobe shown in Figure 7c is about 20 $\mu Jy\,arcsec^{-2}$, while that at 37.1 $\mu m$ is about 170 $mJy\,arcsec^{-2}$. The corresponding power [$\nu F_{\nu}$] of the infrared radiation is almost 100 times that of the visible, suggesting that the scattering grains in the West have very low net albedo. Note that we are explicitly assuming at this point that the starlight absorbed by these grains heats them to produce the radiation seen by \emph{SOFIA}, while that scattered is seen at 0.547 $\mu m$ by HST and that the stellar temperature is about 5000 $K$ as suggested by Table 2. This observation suggests a possible explanation for the apparent decoupling of the infrared and visible light distributions at the Eastern limb. Because of the broad distribution of dust particle sizes in M2-9, it is possible that the scattered light at the Eastern limb comes from an admixture of grains -- perhaps larger and considerably colder than those seen to the West -- that scatter very effectively, increasing the visible brightness with little impact at \emph{SOFIA} wavelengths. It is also conceivable that the marked asymmetry in the scattered light as compared to the high degree of symmetry shown in the infrared images could be due to a foreground absorbing cloud with $\tau_{V}$ $\sim$ 1 which happens to bisect the nebula.  Apart from the improbability of such an alignment, however, there is no evidence for such variable extinction in the HST line emission image shown in Figure 1.
\\

We can further investigate the issue of the low observed optical surface-brightness,
compared to that observed with SOFIA at long wavelengths (e.g., 37.1\,$\mu m$)
using our DUSTY modeling. Our model (described in 5.3.1) gives a 
37.1\,$\mu m$~surface brightness of $S(37.1\,\mu m)=175\,mJy\,arcsec^{-2}$ at a
radial offset of $10{''}$, in good agreement with the observed value, but predicts the optical surface brightness is $S(0.55\,\mu m)=0.65\,mJy\,arcsec^{-2}$, much
{\it higher} than observed. In order to check that this discrepancy is not
simply a problem associated with the optical imaging (e.g., its calibration),
we have examined near-infrared images of M2-9, and find that a discrepancy
exists there as well. 
\\

Hora \& Latter (1994) present ground-based spectroscopy and imaging of M2-9 in
multiple filters in the near-infrared. Their 2.26\,$\mu m$~filter image indicates a total (line+continuum) surface brightness S(2\,$\mu m$)$\sim\,1.5\times10^{-4}\,Jy\,arcsec^{-2}$ at a distance of $10{''}$, and although their spectrum shows that there is weak line emission included within the filter bandpass, there is a weak continuum present as well. We also found an archival near-IR HST image, obtained with NICMOS (NIC2) using the F215N filter on 1998 May 19, as part of GO program 7365 (PI: W. B. Latter). This filter spans 2.14 to 2.16\,$\mu m$, and only has a very weak H$_2$ line within this range. Using the HST pipeline photometry from the image file header, we find  S(2\,$\mu m)\,\sim\,3\times10^{-4}\, Jy\,arcsec^{-2}$, at a radial offset of $10{''}$ offset in the N-lobe. The model-predicted value of S(2\,$\mu m$) is $0.24\times10^{-4}\, Jy\,arcsec^{-2}$, i.e., significantly {\it lower} than either the ground-based or the HST value. 
\\

Hence the observed optical and near-IR surface brightnesses are discrepant from
the model ones, but in opposite directions, suggesting that the radiation
heating the grains in the lobes is redder than the 5000 $K$ of our standard model. This
reddening of the heating radiation can be achieved in two ways (i) assuming a
lower value of $T_{eff}$ for the central star, and (ii) increasing the
extinction in the inner region of the model dust shell. We have computed models
to examine both these effects and find that by lowering $T_{eff}$ from $5000\,K$
to $3000\,K$ and raising $\tau_{V}$ from 1 to 3, we obtain $S(0.55\,\mu m)=35\,
\mu Jy\,arcsec^{-2}$ and 
$S(2\,\mu m)=1.2\times10^{-4}\, Jy\,arcsec^{-2}$, in better agreement with
their observed values. Both of the above changes can be accommodated in our
models of the central source and the lobes. The central disk model is not very
sensitive to the adopted value of $T_{eff}$ of the central star. An increase in
$\tau_{V}$, together with no significant change in the total far-infrared model
fluxes and the input luminosity, can be achieved with a decrease in
the solid angle of the dust shell subtended at the center, by a factor 3 from
its value of 2$\pi$ in our standard model -- such a decrease is not unreasonable (and may
in fact be desirable), given that the lobes in M\,2-9 are very highly
collimated.
In this scenario as well, the increased brightness at the Eastern limb suggests an additional population of larger and colder grains.

\subsubsubsection{Infrared Detection of the Optical Knots}
As illustrated most recently by Co11, the optical images of M2-9 show persistent structures in the form of knots and arcs. Most pronounced are the knots N3 and S3 which lie along the center line about 15${''}$ N and S of the central source (Figure 1). Although their morphologies have varied somewhat with time, these knots persist over the 1999-to-2010 time period sampled by Co11 and can also be seen as far back as the images presented by Allen and Swings (1972). We have searched for these knots by examining scans along the axis of the outflow; the scans at 19.7, 24.2, and 37.1 $\mu m$ are shown in Figure 9. Both N3 and S3 are seen very clearly at 19.7 $\mu m$, approximately equidistant from the central source. The separation of the two knots is about 28${''}$, very close to the 29${''}$ estimated by eye for these somewhat diffuse structures from the Co11 images. The brighter knot, N3, is seen less strongly at 24.2 $\mu m$ than at 19.7 $\mu m$, but neither is seen at 37.1 $\mu m$. Because the knots are seen very strongly in visible emission lines of OIII, NII, and HI, it is tempting to conclude that the enhancement at 19.7 $\mu m$ is due to localized emission from the SIII line at 18.7 $\mu m$, which falls well within the broad 19.7 $\mu m$ filter. This line is very strong in the \emph{Spitzer} spectrum of the Northern Lobe, which included the knot N3 in the slit; there might also be a contribution from an FeII line at 18 $\mu m$. However, the flux 
in the narrow 18.7 $\mu m$ line over the entire $4.7{''}\times11.3{''}$ \emph{Spitzer} slit (Lyk11) is about an order of magnitude less than the excess flux seen over the broad 19.7 $\mu m$ filter in the feature $\sim$ 14${''}$ North of the central source.
Thus another cause, possibly related to a localized population of very small grains at the position of the knot, or perhaps to mechanical heating of the dust grains if the knots are produced by shocks where highly collimated outflows impact slowly moving ambient material, must be sought for the 19.7 $\mu m$ excess. 

\subsection{M2-9 and PN Shaping}
Jet-sculpting models for producing bipolar PNe (e.g., Lee \& Sahai
2003) require a fast, collimated outflow expanding inside a
pre-existing AGB circumstellar envelope (CSE), resulting in
collimated lobes with dense walls, as seen in M2-9 and other bipolar
PNe. 
The true nature of M2-9 has been debated, i.e., whether or not it is a normal PN (i.e., an object in which the optically-visible nebula seen prominently in forbidden line emission, consists of matter ejected by a central star during its AGB phase, that is then photoionized by the same star after it has evolved off the AGB and become much hotter). The current consensus appears to be that M2-9 is not a normal PN, but that its central star is an AGB or young post-AGB star with a hot white-dwarf companion (e.g., Livio \& Soker 2001), and thus represents the ``long-period interacting binary" evolutionary channel for the formation of PN-like nebulae (Frew \& Parker 2010). In either case, one would expect to see the presence of the mass ejected during the AGB phase, in the form of a circumstellar envelope around the bipolar lobes. So it appears somewhat surprising that 
we have seen no direct observational evidence of the AGB CSE in M2-9 so far. However, our inability to detect material outside the lobes in M2-9, e.g., via
scattered light at optical wavelengths, or thermal emission from
dust, or molecular-line emission, may simply be a sensitivity issue,
if the AGB mass-loss rate in M2-9 was relatively low. Note that
the lateral expansion of the lobes is set by the sound speed in
ionized gas ($\sim$ 10 $km/s$ at 10$^4$ K), which is small compared to
the much higher axial speed of the lobes (145 $km/s$, Co11), hence even if the confining pressure of the ambient medium
is relatively low for a low AGB mass-loss rate, the lobes would
maintain their collimated shapes. 

The extensive optical imaging survey of young PNe with HST by Sahai et al. (2011b) found faint halos in a significant fraction, but not all, of their sample. Deep optical imaging of M2-9 with HST would be very useful. Molecular material in the halo may be difficult to detect since photodissociation by the general
interstellar UV field might be significant, especially if the AGB
mass-loss rate was relatively low. 

\subsection{Directions for Further Work}
The main new results of this paper refer to the spatial and particle size distribution of the dust seen in the thermal infrared and at mm-wavelengths.  Along the way, however, we have identified several additional areas where the work to date poses interesting, unanswered questions:
\\

Firstly, the identification of transient heating of small particles as an important contribution to the radiation from the lobes beyond $\sim\,10{''}$ from the central source calls for an analysis of the extended emission which would go beyond the simple DUSTY model reported here and include transient heating and  explore a range of grain materials and sizes. Such modeling could also address the uncertainty in the stellar temperature discussed in 5.3.4; it would also be appropriate to use cylindrical coordinates in this improved analysis.
\\

Secondly, we note that the model for the central compact source given in Table 2 is based on silicate grains rather than the amorphous carbon which we chose to describe the grains in the lobes. This dichotomy is consistent with the fact that the spectra of M2-9 show silicate absorption in the central source and PAH emission in the lobes (Lyk11).
\\

It therefore appears that M2-9 belongs to the well-known subclass of post-AGB objects that have been labeled as having ``mixed-chemistry" (e.g., Morris 1990; Waters et al. 1998a,b; Cohen et al. 1999, 2002). Since, during the AGB phase, a star may evolve from being oxygen rich to being carbon rich (due to the 3rd dredge up), a popular hypothesis for this phenomenon is that the disk formed (e.g., by gravitational capture of the stellar wind around a close companion) when the central star was still oxygen rich, whereas the extended emission is due to a more recent carbon-rich outflow. But a difficulty with applying this hypothesis to M2-9 is that the gaseous nebula in this object is known to be O-rich, with C/O $<$ 0.5 (Liu et al. 2001).
\\

Guzman-Ramirez et al. (2011) propose an alternative hypothesis, based on the strong correlation between the presence of a dense torus and mixed-chemistry in their sample of 40 objects. They argue that the popular hypothesis cannot explain the widespread presence of the mixed-chemistry phenomenon among PNe in the Galactic bulge (Perea-Calderon et al. 2009, Guzman-Ramirez et al. 2011), as these old, low-mass stars should not go through the 3rd dredge-up. They suggest that the mixed-chemistry phenomenon in Galactic bulge planetary nebulae may be due to hydrocarbon chemistry in an UV-irradiated, dense torus that produces long-carbon chain hydrocarbons that then produce PAHs.  Given that PAH features have been observed in the lobes of M2-9, it is plausible that the UV-irradiation hypothesis is responsible for the presence of small carbon-rich grains in the lobes.  We suggest that the new data presented here on the spatial and size distributions of the grains, together with our suggestions concerning large grains in both the lobes and the disk, make M2-9 a detailed astrophysical laboratory for further study of the processes which produce these mixed-chemistry objects.

\section{CONCLUSIONS}
We have presented and analysed images of M2-9 with $\sim$ 4${''}$-to-5${''}$ resolution in six infrared bands at wavelengths between 6.6 and 37.1 $\mu m$. The principal new results from these \emph{SOFIA} observations of M2-9 center around the spatial and size distribution of the grains which produce the infrared radiation from the outflow lobes in this bipolar nebula. The spatial distribution of the emission implies that the lobes are fairly uniformly filled with dust, with a marked increase in the dust density in a relatively narrow cylindrical annulus -- with width order 5\% of the lobe radius -- at the outer edge of the lobes. We caution that the spatial resolution of the \emph{SOFIA} observations, in comparison with the width of the lobes, does not permit the model parameters to be pinned down definitively; for example, we can not determine whether this outer annulus lies within or exterior to the optically visible lobes. However, the result that dust is well mixed over the interior of the lobe is well-established.  The side to side asymmetry seen in the HST continuum image of M2-9 is not seen in the infrared images, although it would have been apparent at \emph{SOFIA}'s resolution. The reason for this puzzling discrepancy is unclear, but it may be related to the range of grain sizes which characterize this source; the M2-9 lobes have comparable masses of particles with radii $> 1$ $\mu m$ and with radii $< 0.1$ $\mu m$.
\\

The other principal results of this work are:
\\

\noindent
1. At wavelengths from 6.6 to 37.1 $\mu m$, the image of M2-9 is dominated by a bright central point source, which is not definitely resolved at any wavelength with \emph{SOFIA}'s $\sim$ 4-to-5${''}$ beams. 
\\

\noindent
2. The extended bipolar lobes are clearly seen at wavelengths from 19.7 to 37.1 $\mu m$; the integrated emission from the lobes is detected down to 6.6 $\mu m$.
\\

\noindent	
3. The infrared emission of the point source at wavelengths between 11.1 and 24.2 $\mu m$ agrees extremely well with the predictions of a disk model based on Lyk11. The fit suggests a distance of 1200 $pc$ to M2-9 and a luminosity of 2500 $L_{\odot}$ for the star which powers it, although the distance and hence the luminosity are not well-constrained. Emission from the point source in excess of the model is seen shortward of 11.1 $\mu m$, likely contributed by warm dust close to the star that resides in a different geometrical component than the disk, such as the inner part of a disk wind. Assuming isotropic emission, we find that the total 2.5-40 $\mu m$ infrared luminosity of the compact source is $\sim$ 840 $L_{\odot}$, and that of the lobes $\sim$ 390 $L_{\odot}$. The total observed 2.5-120 $\mu m$ luminosity of M2-9, including the $IRAS$ measurements at longer wavelengths, and assuming isotropic emission, is $\sim$ 1530 $L_{\odot}$. Because the emission is clearly not isotropic, this is not inconsistent with the 2500 $L_{\odot}$ inferred from the disk model. 
The \emph{SOFIA} photometry agrees well with that obtained from other platforms, including $ISO$, $WISE$ and $IRAS$. 
\\

This work shows that compact planetary nebulae are ideal targets for study from \emph{SOFIA}, not only photometrically but with other capabilities, most notably grism spectroscopy, now becoming available on this new airborne observatory.
\\

We thank the staff and crew of the \emph{SOFIA} observatory for their support in carrying out these observations at an epoch when \emph{SOFIA} was still in its development phase. 
\\

We thank Bruce Balick for encouragement and useful discussions and comments, and the referee for a very useful report. Portions of the work were carried out at the Jet Propulsion Laboratory operated by the California Institute of Technology under a contract with NASA.
\\

J. Davis was the Charles and Valerie Elachi SURF Fellow, under Caltech's Summer Undergraduate Research Fellow program, while working on this project.
\\

This work is based on observations made with the NASA/DLR Stratospheric Observatory for Infrared Astronomy (\emph{SOFIA}). \emph{SOFIA} is jointly operated by the Universities Space Research Association, Inc. (USRA), under NASA contract NAS2-97001, and the Deutsches SOFIA Institut (DSI) under DLR contract 50 OK 0901 to the University of Stuttgart. This research used the Hubble Legacy Archive (HLA) which is a collaboration between the Space Telescope Science Institute (STScI/NASA), the Space Telescope European Coordinating Facility (ST-ECF/ESA) and the Canadian Astronomy Data Centre (CADC/NRC/CSA).

\begin{deluxetable}{cccccc}
\tablecaption{M2-9 Photometry and Size Data}	
\tablehead{
\hline
\colhead{Wavelength}	&	\multicolumn{3}{|c|}{Flux} 	& 	\multicolumn{1}{|c|}{FWHM}		&	\multicolumn{1}{|c|}{Beam Size} 	\\
\hline
			$\mu$m		&	\multicolumn{3}{|c|}{Jy}		&	\multicolumn{1}{|c|}{arcsec}			&	\multicolumn{1}{|c|}{arcsec}			\\
\hline
					&	\multicolumn{1}{|c|}{Point Source}	&	\multicolumn{1}{|c|}{Total Flux}	& \multicolumn{1}{|c|}{Extended Component}	& &}
\startdata
3	&	4.6	&		&		&		&		\\
3.7	&	7.9	&		&		&		&		\\
4.5	&	15	&		&		&		&		\\
6.6	&	24	&	30.2	&	6.2	&	3.7	&	3.68	\\
11.1	&	32	&	46.9	&	$14.9\pm5.4*$	&	3.7	&	3.85	\\
19.7	&	58	&	87.5	&	29.5	&	3.9	&	3.76	\\
24.2	&	55	&	93.4	&	38.9	&	4.1	&	4.19	\\
33.6	&	63	&	157.7	&	94.7	&	4.5	&	4.42	\\
37.1	&	48	&	138.4	&	90.4	&	4.9	&	4.51	\\
12	&		&	50	&		&		&		\\
25	&		&	110	&		&		&		\\
60	&		&	123	&		&		&		\\
100	&		&	76	&		&		&		\\
1300	&	0.21	&	0.36	&	0.15	&		&		\\
\enddata
\tablecomments{Infrared photometry of M2-9. Data from 6.6 to 37.1 $\mu$m, is from this paper. 3-4.5 $\mu$m data is from Lyk11. The central source is unresolved at these wavelengths by \emph{ISO}'s 1${''}$ to 2${''}$ PSF. Also included is 12-100 $\mu$m data from the \emph{IRAS} Point Source Catalog (beam size $\gtrsim$ 1${'}$) and the 1.3 $mm$ measurements of SC98.  The right two columns compare the FWHM of the compact central source in our \emph{SOFIA} images with the beam size determined for our flight series by the \emph{SOFIA} Science Center. The quoted ($2\sigma$) uncertainties in this beam size are $>10$\% at all wavelengths. *Error determined from variations in brightness of reference positions. For all other \emph{SOFIA} measurements, the statistical errors are smaller than the $\sim$ 20\% calibration uncertainty.}
\end{deluxetable}

\begin{deluxetable}{ l  l  l }
\tablecaption{M2-9 -- Best-Fit Parameters for Central Source Model}
\tablehead{
\hline
\colhead{Parameter}	&	Value 	& 	Comment}
\startdata

  Disk inner radius	&	$15\pm1$ $au$				&	  	\\
  Disk outer radius	&	$800\pm100$ $au$			&		Not well constrained\\
  Mass of dust		&	$1\pm0.1\times10^{-5}$ $M_{\odot}$		&	Draine and Lee astronomical silicates, sizes 0.01-to-5 $\mu m$. \\
					&							&	 (Mathis et al. 1977 size distribution)	\\
  $\alpha$			&	$2.2\pm0.05$	&	Density within disk falls radially as $r^{-\alpha}$	\\
  $\beta$	&	$1.23\pm0.02$	&	Scale height varies radially as $r^{\beta}$.  Disk is flared	\\
  $h_{100}$	&	$37\pm3$ $au$	&	Fiducial scale height at 100 $au$	\\
  $T_{eff}$	&	5000 $K$	&	 Temperature of central star doing most of the heating	\\
  $L$	&	2500 $L_{\odot}$	&	Luminosity of central star	\\
\enddata
\tablecomments{The disk model assumed here can be described by the following density law, described further in Lyk11, where $r$ is the radial and $z$ the vertical coordinate in a cylindrical coordinate system centered on the star, and $R_*$ is the radius of the star:
$\rho(r,z) = \rho_0 (\frac{R_*}{r})^{\alpha} exp(-\frac{1}{2} (\frac{z}{h(r)})^2 )$;\,\, $h(r) = h_0 (\frac{r}{R_*})^{\beta}$
}
\end{deluxetable}														

\clearpage
\begin{figure}
\includegraphics[width=0.99\textwidth]{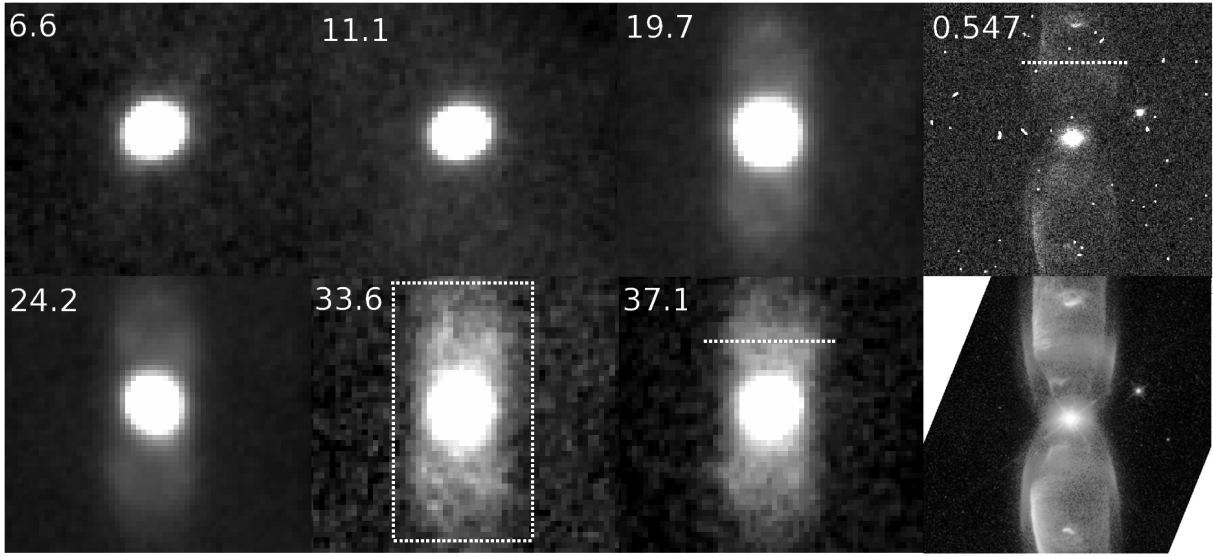}
\caption{Images of M2-9 at six \emph{SOFIA} wavelengths, augmented with data from \emph{HST} in the two right hand panels.  The images are oriented with N to the top and E to the left.  The rectangle on the 33.6 $\mu m$ image is 20${''}\times40{''}$ in size.  The dotted lines 10${''}$ to the N of the central point source trace the path along which the simulations described in the text were calculated.  The HST data, 0.547 $nm$ in the upper right, and a multiband optical emission line image in the lower right, were taken in 1997.}
\end{figure}

\clearpage
\begin{figure}
\includegraphics[width=0.99\textwidth]{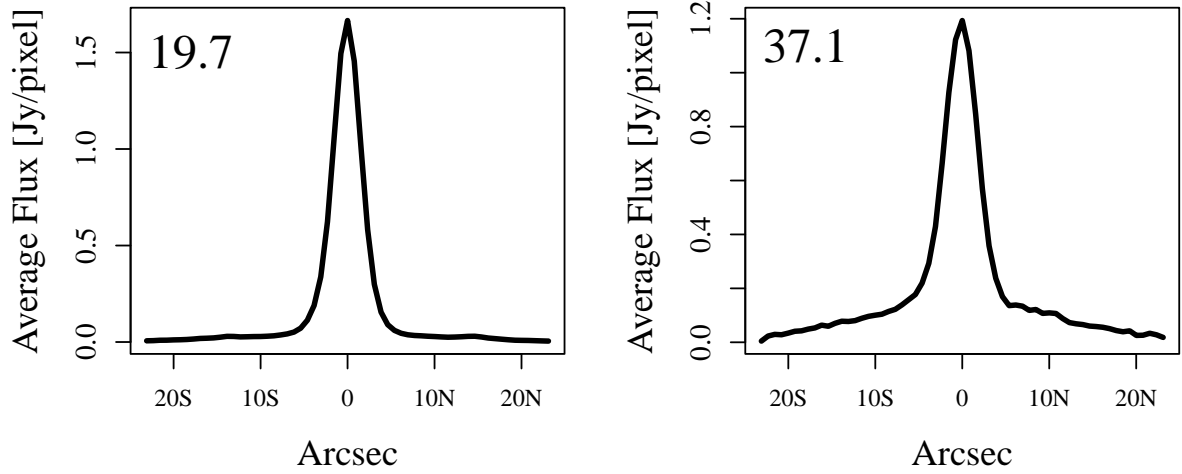}
\caption{N-S scans through the M2-9 point source at 19.7 [left] and 37.1 $\mu m$.  The data are averaged over 5 pixels ($\sim$ 3.8${''}$) in the E-W direction at each wavelength.}
\end{figure}

\clearpage
\begin{figure}
\includegraphics[width=0.99\textwidth]{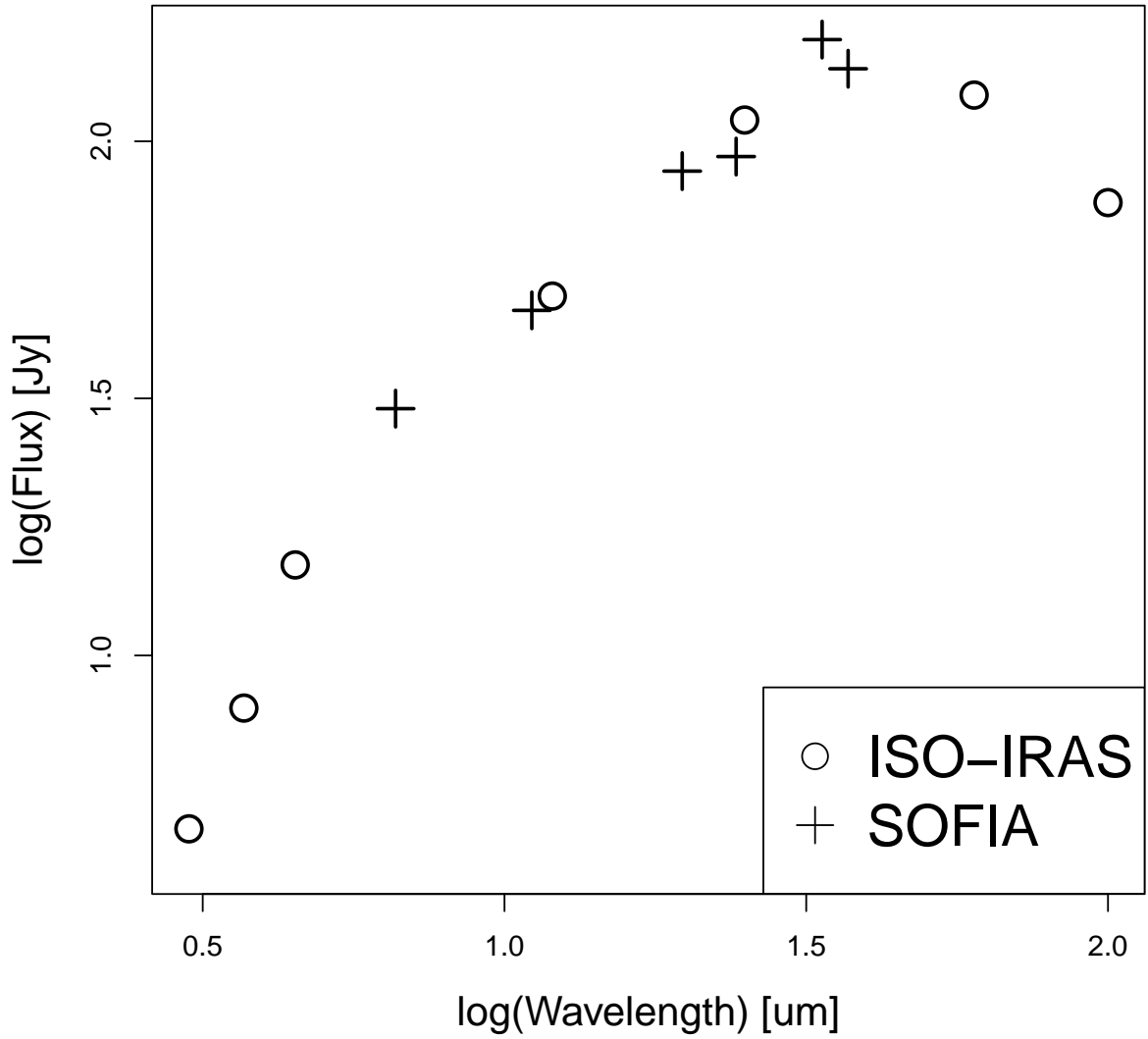}
\caption{2.5-120 $\mu m$ SED of M2-9 including data from \emph{ISO}, \emph{IRAS}, and \emph{SOFIA}.}
\end{figure}

\clearpage
\begin{figure}
\includegraphics[width=0.99\textwidth]{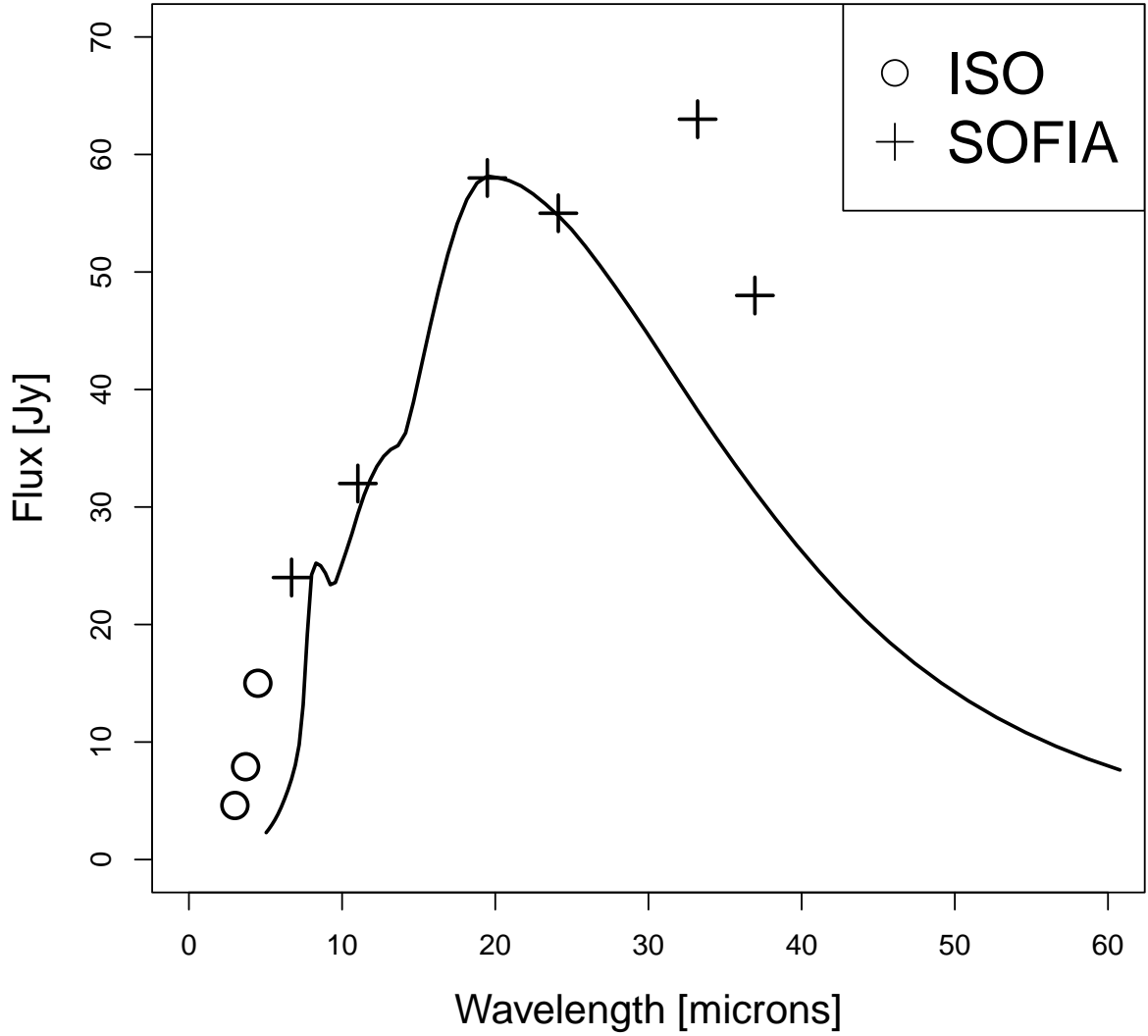}
\caption{\emph{SOFIA} and \emph{ISO} photometry of the central point source of M2-9 compared with the predictions of the model described in Table 2.}
\end{figure}

\clearpage
\begin{figure}
\includegraphics[width=0.99\textwidth]{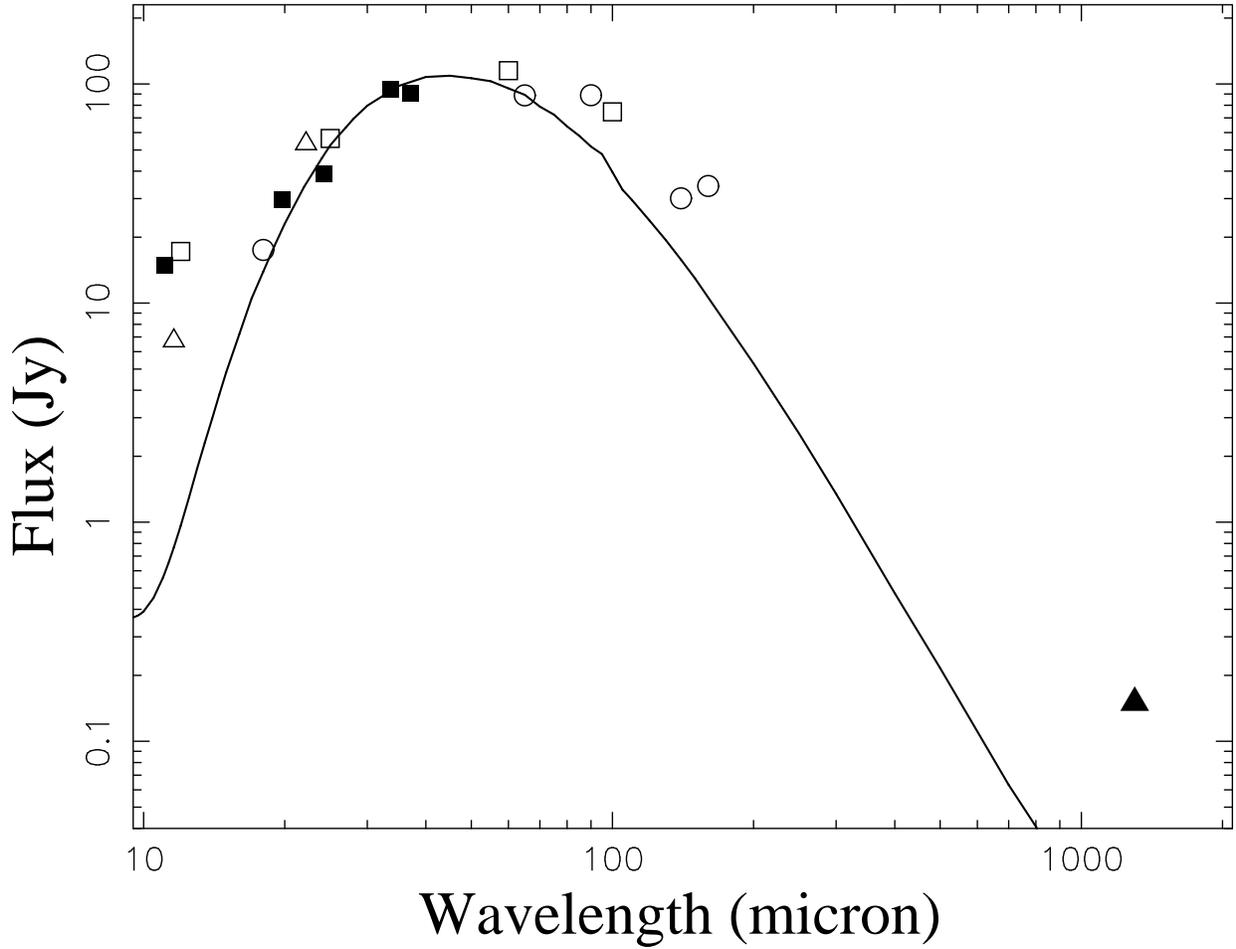}
\caption{Observed (symbols) and model (curve) SED of the M2-9 lobes. The photometric data shown are as follows: filled squares -- \emph{SOFIA}, open squares -- \emph{IRAS},  open triangles -- \emph{WISE}, open circles -- \emph{AKARI}, filled triangle -- \emph{IRAM} 30m. In all cases, the photometry has been corrected for the point source contribution based on the model shown in Figure 4.}
\end{figure}

\clearpage
\begin{figure}
\includegraphics[width=0.99\textwidth]{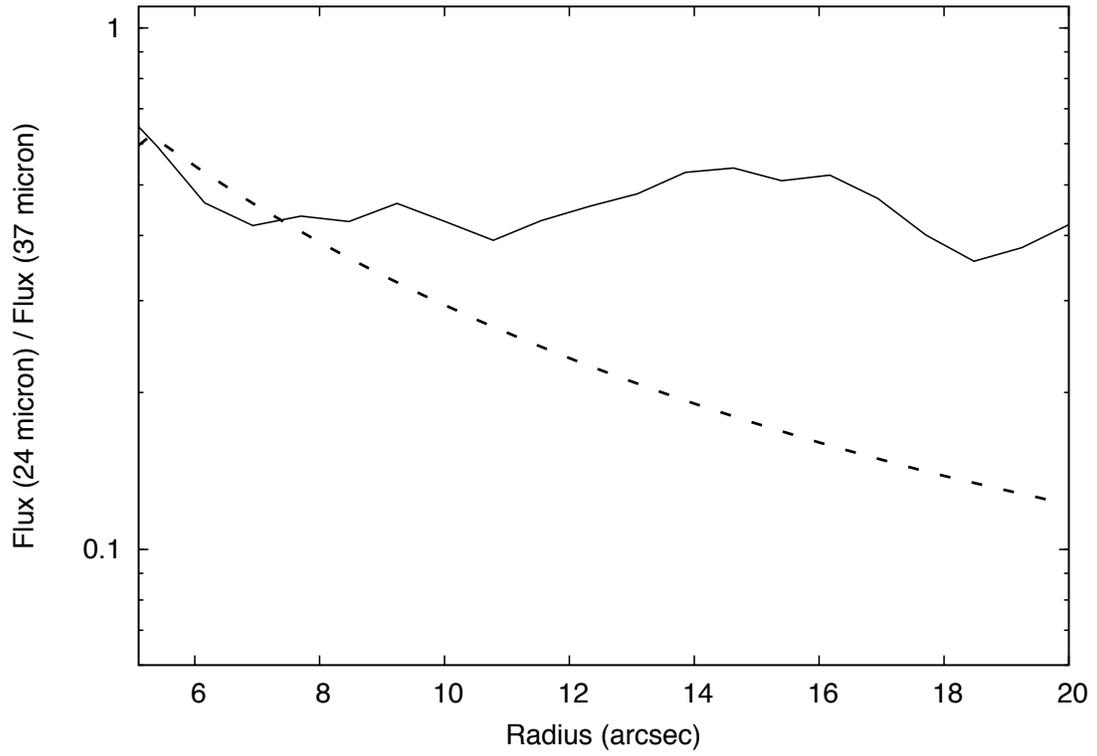}
\caption{24.2-to-37.1 $\mu m$ ratio as a function of position N of the central point source.  Solid line -- Data, averaged over 11 pixels in the E-W direction.  Dotted line -- Prediction of the DUSTY model shown in Figure 5.}
\end{figure}

\clearpage
\begin{figure}
\includegraphics[width=0.99\textwidth]{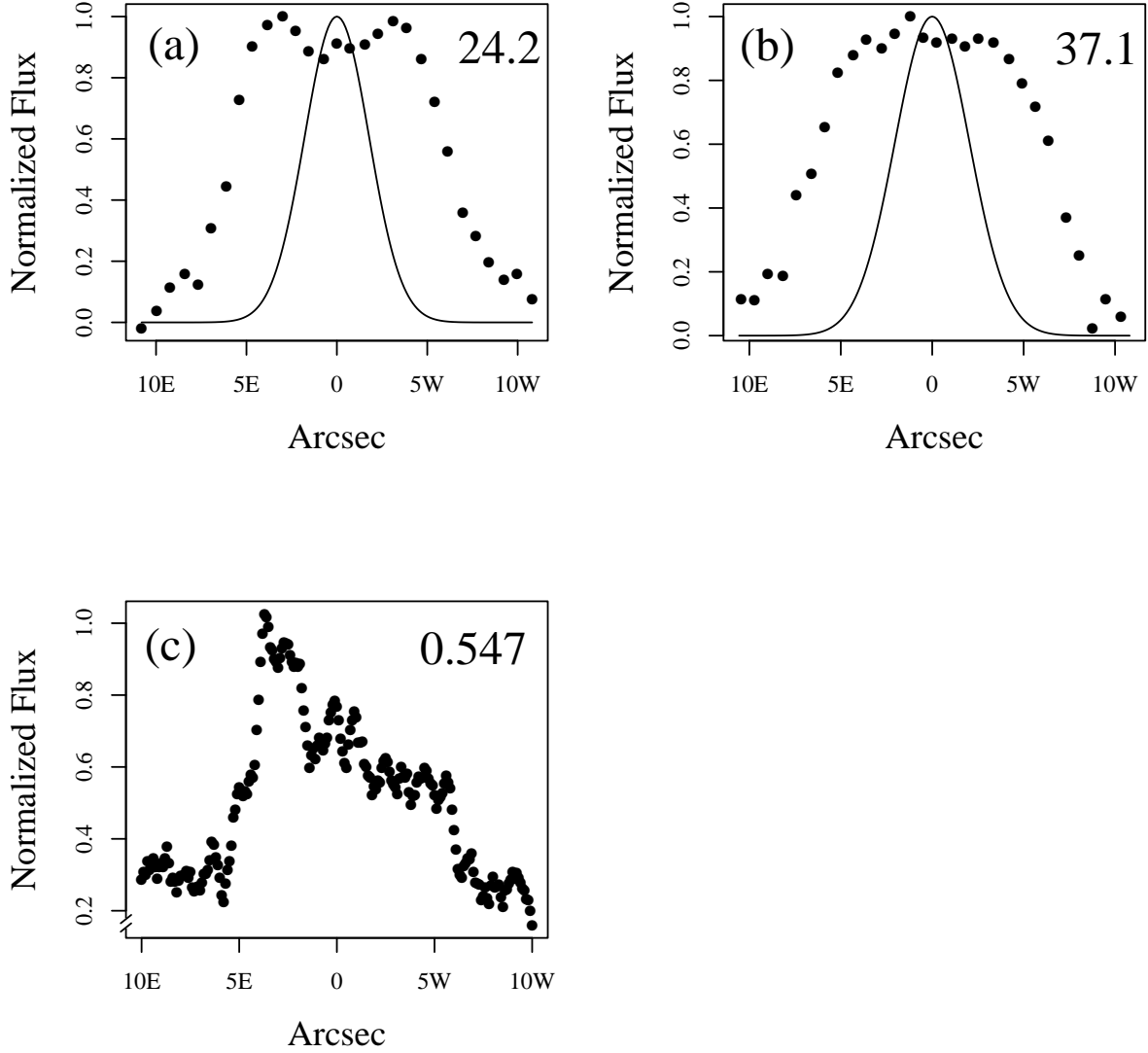}
\caption{Normalized E-W scans 10${''}$ N of the point source through the images of M2-9 at 24.2 (a) and 37.1 (b) $\mu m$. Overlaid on each \emph{SOFIA} scan is the beam profile at that wavelength. For comparison, a similar scan through the 547 $nm$ HST image is shown in panel (c). The 37.1 $\mu m$ data is averaged over three N-S pixels, and the HST data over 20 N-S pixels (the HST pixels are $\sim$ $0.1{''}\times 0.1{''}$ in size). The relative uncertainty in the \emph{SOFIA} data is just a few percent, as shown by the small point-to-point scatter.}
\end{figure}

\clearpage
\begin{figure}
\includegraphics[width=0.99\textwidth]{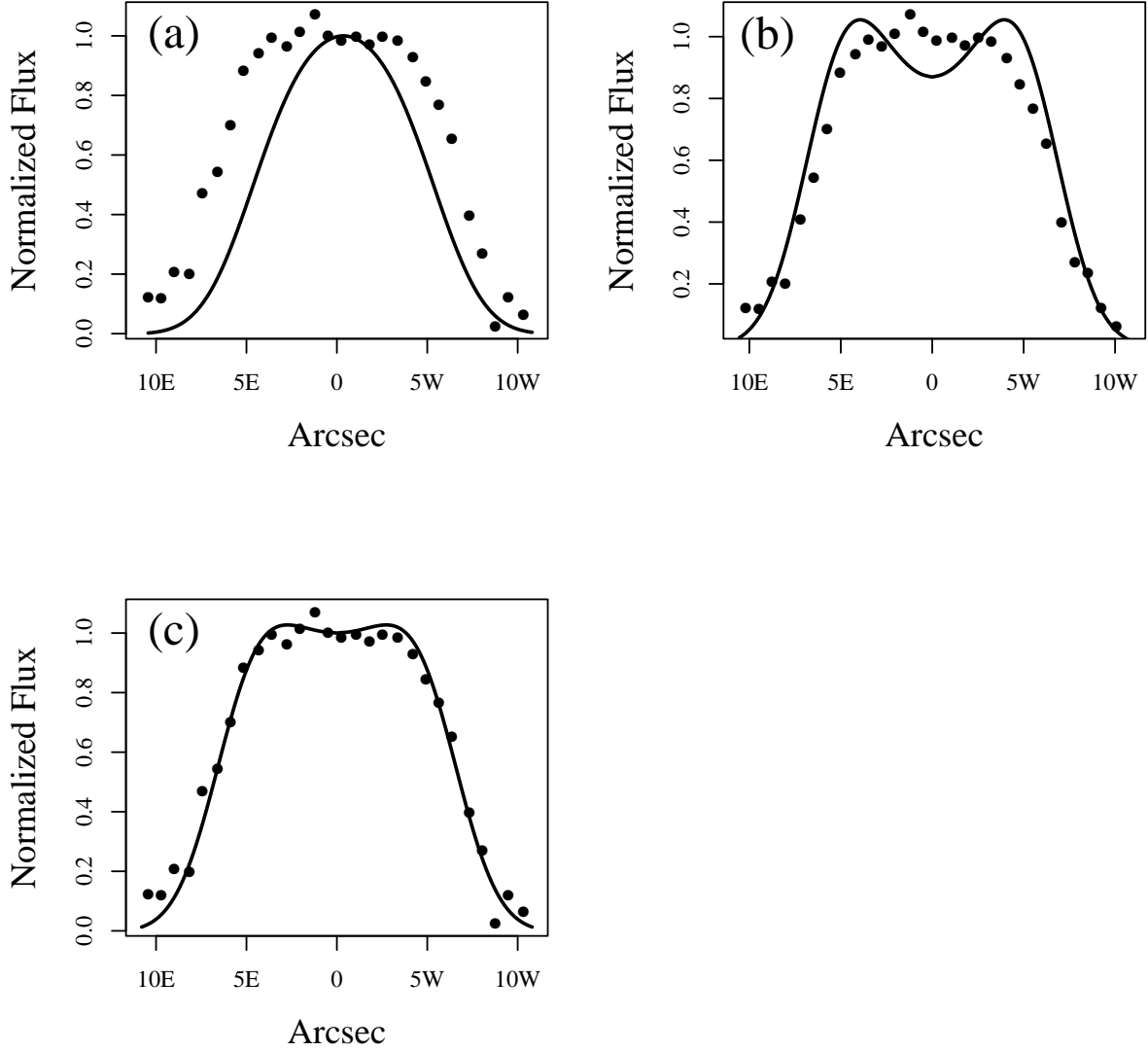}
\caption{Three models for the dust distribution in the Northern lobe are compared with the data  at 37.1 $\mu m$.  Panel (a): a uniform dust distribution which fills the entire lobe.  Panel (b): the dust is confined radially to the outer 5\% of the optically visible lobe.  Panel (c):  30\% of the dust is in the uniform distribution and 70\% is confined to the outer 5\% of the lobe. Note that what is actually modeled is the volume emissivity, which should be very close to the dust distribution as described in the text. }
\end{figure}

\clearpage
\begin{figure}
\includegraphics[width=0.99\textwidth]{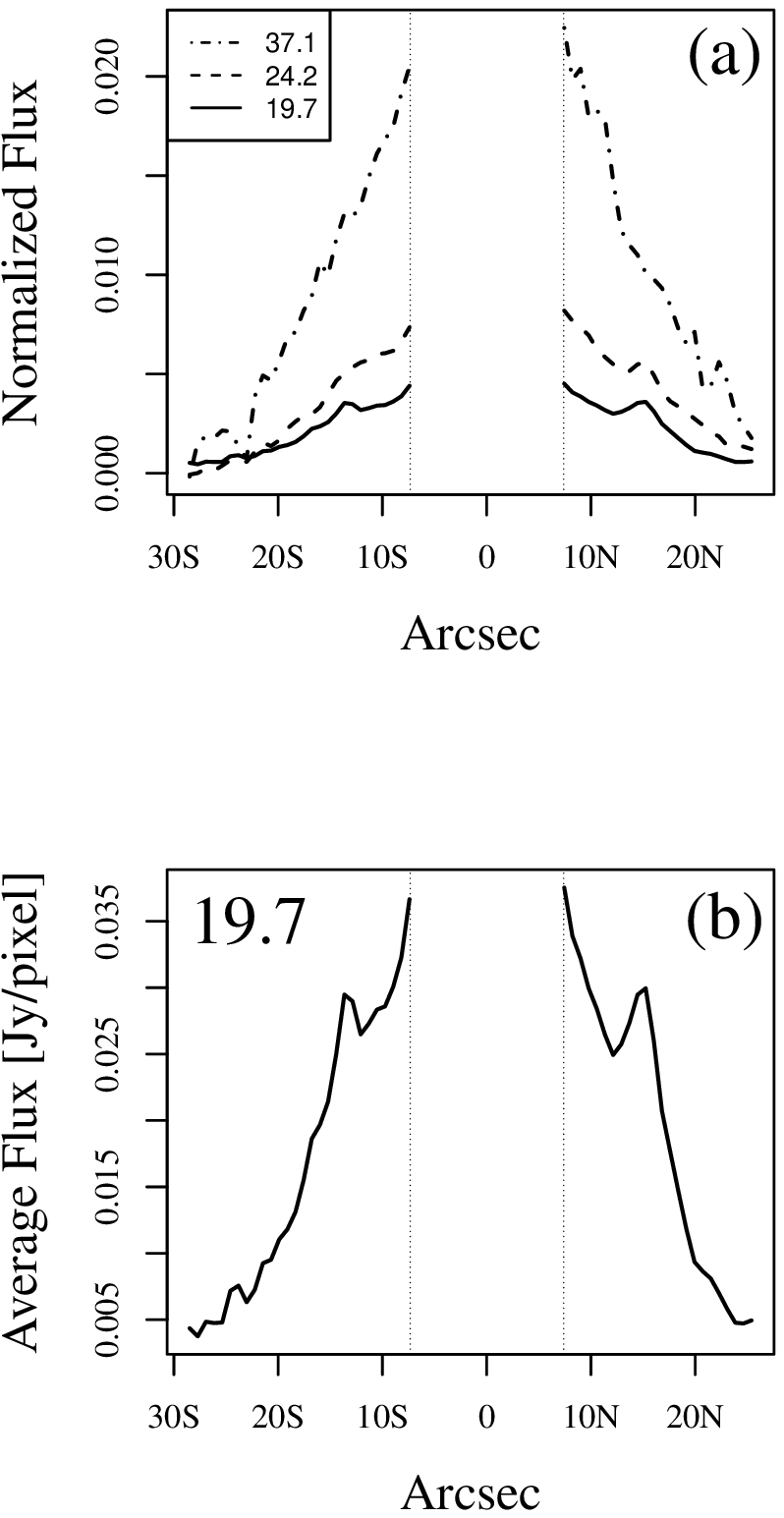}
\caption{
N-S scans along the axis of the M2-9 lobes [cf. Figure 2] are displayed on an expanded scale. In all cases the data are averaged over 5 pixels in the E-W direction. Panel a. (top) shows that the knots $\sim$ 15 arcsec N and S of the central point source are detected at 19.7 $\mu m$ but not at 37.1 $\mu m$; only the northern knot is detected at 24.2 $\mu m$. Panel b. (bottom) shows the 19.7 $\mu m$ scan in units of $Jy/pixel$ to illustrate the absolute brightness of the emission.
}
\end{figure}


\begin{thebibliography}{}
\bibitem[Allen et al.(1972)]{} Allen, D.A., \& Swings, J.P. 1972, ApJ, 174, 583.
\bibitem[Balick et al.(2002)]{} Balick, B., \& Frank, A. 2002, ARA\&A, 40, 439.
\bibitem[Castelaz et al.(1987)]{} Castelaz, M., Sellgren, K., \& Werner, M.W. 1987, ApJ, 313, 853.
\bibitem[Castro-Carrizo et al.(2012)]{} Castro-Carrizo, A., Neri, R., Bujarrabal, V., Chesneau, O., Cox, P., \& Bachiller, R. 2012, A\&A, 545, 1.
%
\bibitem[Chesneau et al.(2007)]{} Chesneau, O., Lykou, F., Balick, B., Lagadec, E., Matsuura, M., Smith, N., Spang, A., Wolf, S., \& Ziljstra, A.A. 2007, A\&A, 473, 29.
\bibitem[Cohen et al.(1999)]{} Cohen, M., Barlow, M. J., Sylvester, R. J., Liu, X.-W., Cox, P., Lim, T., Schmitt, B., \& Speck, A. K. 1999. ApJ. 513. L135.
\bibitem[Cohen et al.(2002)]{} Cohen, M., Barlow, M. J., Liu, X.-W., \& Jones, A. F. 2002, MNRAS, 332, 879.
\bibitem[Corradi et al.(2011)]{} Corradi, R.L.M., Balick, B., \& Santander-Garcia, M. 2011, A\&A, 529, 43.
\bibitem[Doyle et al.(2000)]{} Doyle, S., Balick, B., Corradi, R.L.M., \& Schwarz, H.E. 2000, AJ, 119, 1339.
\bibitem[Frew et al.(2010)]{} Frew, D.J., \& Parker, Q.A. 2010, P.A.P.A, 27, 129.
\bibitem[Garcia-Segura et al.(1997)]{} Garcia-Segura G. 1997. Ap. J. Lett. 489, L189.
\bibitem[Guzman-Ramirez et al.(2011)]{} Guzman-Ramirez, L., Zijlstra, A. A., Nichuimin, R., Gesicki, K., Lagadec, E., Millar, T. J., \& Woods, P. M. 2011, MNRAS, 414, 1667.
\bibitem[Herter et al.(2012)]{} Herter, T.L., Adams. J., De Buizer, J.M., Gull, G.E., Schoenwald, J., Henderson, C.P. Keller, L.D., et al. 2012, ApJ, 749, L18.
\bibitem[Jura et al.(1986)]{} Jura, M. 1986, ApJ, 303, 327.
\bibitem[Jura et al.(2001)]{} Jura, M., Webb, R.A., \& Kahane, C. 2001. ApJ, 550, L71.
\bibitem[Kwok et al.(1985)]{} Kwok, S., Purton, C.R., \& Spoelstra, T.A.T. 1985, A\&A, 144, 321.
\bibitem[Lagadec et al.(2011)]{} Lagadec, E., Verhoelst, T., Merkarnia, D., Suzeea, O., Zijlstra, A.A., Bendoya, P., Szczerba, R., et al. 2011, MNRAS, 417, 32.
\bibitem[Lee et al.(2003)]{} Lee, C.-F., \& Sahai, R. 2003, ApJ, 586, 319.
\bibitem[Livio et al.(2001)]{} Livio, M., \& Soker, N. 2001, ApJ, 552, 685.
\bibitem[Liu et al.(2011)]{} Liu, X.-W., Barlow, M. J., Cohen, M., Danziger, I. J., Luo, S.-G., Baluteau, J. P., Cox, P., Emery, R. J., Lim, T., \& Pequignot, D. 2011, MNRAS, 323, 343.
\bibitem[Lykou et al.(2011)]{} Lykou, F., Chesneau, O., Zijlstra, A.A., Castro-Carrizo, A., Lagadec, E., Balick, B. \& Smith, N.2011, A\&A, 527, L105.
\bibitem[Mathis et al.(1977)]{} Mathis, J. S., Rumpl, W., Nordsieck, K. H. 1977, ApJ, 217, 425.
\bibitem[Matt et al.(2006)]{} Matt, S., Frank, A., \& Blackman, E.G. 2006, Ap. J. Lett, 647, L45.
\bibitem[Morris et al.(1987)]{} Morris, M. 1987, PASP, 99, 1115.
\bibitem[Morris et al.(1990)]{} Morris, M. 1990, ``From Miras to Planetary Nebulae: Which Path for Stellar Evolution?'', eds: M.O. Mennessier \& A. Omont, Editions Frontieres, page 520.
\bibitem[Perea-Calderon et al.(2009)]{} Perea-Calderon, J. V., Garcia-Hernandez, D. A., Garcia-Lario, P., Szczerba, R., \& Bobrowsky, M. 2009, A\&A, 495, L5.
\bibitem[Sahai et al.(2011a)]{} Sahai, R., Claussen, M.J., Schnee, S., Morris, M.R., \& Sanchez-Contreras, C. 2011, ApJ, 739, L3. (Sahai et al. 2011a)
\bibitem[Sahai et al.(2011b)]{} Sahai, R., Morris, M.R., \& Villar, G.G. 2011, AJ. 141, 134. (Sahai et al. 2011b)
\bibitem[Sahai et al.(1998)]{} Sahai, R. \& Trauger, J.T., 1998. AJ, 116, 1357.
\bibitem[Sanchez-Contreras et al.(1998)]{} Sanchez-Contreras, C., Alcolea, J, Bujarrabal, V., \& Neri, R. 1998, A\&A, 337, 233.
\bibitem[Sarkar et al.(2006)]{} Sarkar, G., \& Raghvendra, S. 2006, ApJ, 644, 1171.
\bibitem[Schmeja et al.(2001)]{} Schmeja, S., \& Kimeswerner, S., 2001. A\&A, 377, L18.
\bibitem[Schwarz et al.(1992)]{} Schwarz, H.E., Corradi, R.L.M., \& Melnick, J. 1992, A\&AS, 96, 23.
\bibitem[Smith et al.(2005)]{} Smith, N., \& Gehrz, R. 2005, AJ, 129, 969.
\bibitem[Solf et al.(2000)]{} Solf, J. 2000, A\&A, 354, 674.
\bibitem[Waters et al.(1998a)]{} Waters, L. B. F. M., Beintema, D. A., Zijlstra, A. A., de Koter, A., Molster, F. J., Bouwman, J., de Jong, T., Pottasch, S. R., \& de Graauw, Th. 1998 A\&A, 331, L61.
\bibitem[Waters et al.(1998b)]{} Waters, L. B. F. M., Waelkens, C., Van Winckel, H., Molster, F. J., Tielens, A. G. G. M., van Loon, J. Th., Morris, P. W., Cami, J., Bouwman, J., de Koter, A., de Jong, T., \& de Graauw, Th. 1998, Nature, 391, 868.
\bibitem[Zuckerman et al.(1986)]{} Zuckerman, B., \& Aller, L.H. 1986, ApJ. 301.772.
\bibitem[Zweigle et al.(1997)]{} Zweigle, J. Neri, R., Bachiller, R., Bujarrabal, V., \& Grewing, M. 1997, A\&A, 324, 624.
\end{thebibliography}
\end{document}